\documentclass[preprint,superscriptaddress]{revtex4-1}

\usepackage{graphicx}
\usepackage{amsmath,amssymb,amsfonts}
\usepackage{pgf}
\usepackage{tikz}
\usetikzlibrary{arrows,automata}
\usepackage{color,xcolor}
\usepackage{hyperref}

\newcommand{\agul}{\ensuremath{\alpha_{\rm gu}}}
\newcommand{\ask}{\ensuremath{\alpha_{\rm sk}}}

\newcommand{\turin}{Computer Science Department, University of Turin, Italy}
\newcommand{\sice}{School of Informatics, Computing, and Engineering, Indiana University, Bloomington, IN, USA}
\newcommand{\iuni}{Network Science Institute, Indiana University, IN, USA}
\newcommand{\Army}{U.S Army Research Laboratory, 2800 Powder Mill Rd., Adelphi, MD 20783 USA.}
\newcommand{\RPI}{Network Science and Technology Center, Rensselaer Polytechnic Institute, 335 Materials Research Center 110 8th St. Troy, NY 12180 USA}

\begin{document}

\title{Network Segregation in a Model of Misinformation and Fact-checking}

\author{Marcella Tambuscio}
\email[Corresponding author: ]{marcella.tambuscio@gmail.com}
\affiliation{\turin}

\author{Diego F.M. Oliveira}
\affiliation{\sice}
\affiliation{\Army} 
\affiliation{\RPI}

\author{Giovanni Luca Ciampaglia}
\affiliation{\iuni}

\author{Giancarlo Ruffo}
\affiliation{\turin}

\date{\today}

\begin{abstract}
Misinformation under the form of rumor, hoaxes, and conspiracy theories spreads on social media at alarming rates. One hypothesis is that, since social media are shaped by homophily, belief in misinformation may be more likely to thrive on those social circles that are segregated from the rest of the network. One possible antidote to misinformation is fact checking which, however, does not always stop rumors from spreading further, owing to selective exposure and our limited attention. What are the conditions under which factual verification are effective at containing the spreading of misinformation? Here we take into account the combination of selective exposure due to network segregation, forgetting (i.e., finite memory), and fact-checking. We consider a compartmental model of two interacting epidemic processes over a network that is segregated between gullible and skeptic users. Extensive simulation and mean-field analysis show that a more segregated network facilitates the spread of a hoax only at low forgetting rates, but has no effect when agents forget at faster rates. This finding may inform the development of mitigation techniques and raise awareness on the risks of uncontrolled misinformation online.    
\end{abstract}

\maketitle

\section{Introduction}

Social media are rife with inaccurate information of all sorts~\cite{Bakshy2011,friggeri2014rumor,DelVicario2016}. This is in part due to their egalitarian, bottom-up model of information consumption and production~\cite{benkler2006wealth}, according to which users can broadcast to their peers information vetted by neither experts nor journalists, and thus potentially inaccurate or misleading~\cite{Kwak:2010:TSN:1772690.1772751}. Examples of social media misinformation include rumors~\cite{friggeri2014rumor}, hoaxes~\cite{Nyhan2013}, and conspiracy theories~\cite{anagnostopoulos2014viral,galam2003modelling}. 

In journalism, corrections, verification, and fact-checking are simple yet powerful antidotes to misinformation~\cite{borel2016chicago}, and several newsrooms employ these techniques to vet the information they publish. Moreover, in recent years, several independent fact-checking organizations have emerged with the goal of debunking widely circulating claims online. From now on, we refer to all these practices collectively as \emph{fact-checking}. Among the leading US-based fact-checking organizations we can cite Snopes~\cite{Snopes.com}, FactCheck.org~\cite{Factcheck.org}, and Politifact~\cite{Politifact.com}. Several more are joining their ranks worldwide~\footnote{The Duke Reporters' Lab keeps an updated list of global fact-checking sites at \url{https://reporterslab.org/fact-checking/}.}. In many cases these organizations cannot cope with the sheer volume of misinformation circulating online, and some are exploring alternatives to scale their verification efforts, including automated techniques~\cite{Ciampaglia2015}, and collaboration with technology platforms such as Facebook~\cite{Owens2015} and Google~\cite{dong2015knowledge}. 

These trends thus beg a rather fundamental question --- is the dissemination of fact-checking information effective at stopping misinformation from spreading on social media? In particular cases, timely corrections are enough to limit a rumor from spreading further~\cite{Andrews:2016:KUT:2818048.2819986,friggeri2014rumor,nyhan2015effect}. However, administering fact-checking information may also have adverse effects. For example, in some instances it has been observed that correcting an inaccurate or misleading claim can have counterproductive effects, increasing --- and not decreasing --- belief in it. This is a phenomenon called the \emph{backfire effect}~\cite{Nyhan2013}. Recent work has however failed to replicate this form of backfiring in independent trials, suggesting that it is a rather elusive phenomenon~\cite{wood2016elusive}. 

Fact-checking could also lead to a \emph{hypercorrection effect}, meaning that providing accurate information to people who have been exposed to misinformation may cause them, on the long term, to forget the former, and remember the latter~\cite{butler2011hypercorrection}. Thus, given the growing emphasis put into fact-checking, as well as its unintended side effects, it is clear that, for a better understanding of how to fight social media misinformation, it would be useful to explore the relation between fact-checking and the misinformation it is intended to quell. 

Recent work has also revealed that, when it comes to misinformation, online conversations tend to be highly polarized~\cite{Bessi2015,DelVicario2016}. This suggests the importance of homophily and segregation in the spread of misinformation. 
Since social networks are shaped by homophily~\cite{McPherson2001}, one hypothesis is that misinformation may be more likely to thrive in those social circles that are \emph{segregated} from the rest of the network. Social media may be particularly susceptible to this aspect due to the fact that exposure to information is mediated in part by algorithms, whose goal is to filter and recommend stories that have a high potential for engagement. This may create filter bubbles and echo chambers, information spaces that favor confirmation bias and repetition~\cite{Pariser2011,sunstein2002republiccom}. Recent work has started to measure the extent to which editorial decisions performed automatically by algorithms affect selective exposure, and thus segregation of the information space~\cite{Bakshy2015,nikolov2015measuring}. Therefore, in modeling the interplay between misinformation and fact-checking, our second goal is to shed light on the role of the underlying social network structure in the spreading process, in particular the presence of communities of users with different attitude toward unvetted and unconfirmed information --- which could potentially constitute misinformation. 

Besides segregation, in the literature there is also disagreement about whether \emph{weak ties} --- the links that connect different communities together --- play a role in the diffusion of information. Some studies suggest that weak ties play an important role~\cite{bakshy2012role}; others that they do not~\cite{onnela2007structure}. In their seminal work on complex social contagion, Centola and Macy argue that the spread of collective actions benefits from \emph{bridges}, i.e. ties that are ``wide enough to transmit strong social reinforcement''~\cite{centola2007complex}. 
It is well known that misinformation can be propagated thanks to \emph{repetition}~\cite{knapp1944psychology,allport1947psychology}, which in some ways can be obtained through social reinforcement, and thus it would be useful to investigate this additional aspect as well.

In terms of modeling, there has hitherto been little work on characterizing the epidemic spreading of different types of information, with most efforts devoted to describing mutually independent processes~\cite{funk2010interacting,newman2013interacting}. Instead, the presence of the rich cognitive effects just described suggests that misinformation and fact-checking interact and compete for the attention of individuals on social media, and this could lead to non-trivial diffusion dynamics. Among the work specifically devoted to competition in the diffusion of information, or memes, the literature has focused on the role of limited attention~\cite{weng2012competition,gleeson2014competition}, as well as that of information quality~\cite{qiu2017limited,nematzadeh2017how}. 

Several models have been proposed in prior work to describe the propagation of rumors in a complex social networks ~\cite{daley1964epidemics,chierichetti2009rumor,moreno2004dynamics,acemoglu2010spread}. Most are based on the epidemic compartmental models like the SIR (Susceptible--Infected--Recovered)~\cite{Pastor-Satorras2015}, contemplating the fact-checking only as a remedy \emph{after} the hoax infection. Another class of models uses branching processes on signed networks to take into account user polarization~\cite{DelVicario2016}. Neither type, however, takes into account in the same model the three aforementioned mechanisms --- competition between hoax and fact-checking, forgetting mechanisms and segregation. 

To consider all these features, here we introduce a simple agent-based model in which individuals are endowed with finite memory and a fixed predisposition toward factual verification. In this model, hoax and fact checks compete on a network formed by two groups, the \emph{gullible} and the \emph{skeptic}, marked by a different tendency to believe in the hoax. Varying the level of segregation in the network, as well as the relative credibility of the hoax among the two groups, we look at whether the hoax becomes endemic or instead is eradicated from the whole population. 

\section{Model} 

\begin{figure}[tbp]
    \centering
	\begin{tikzpicture}[->,>=stealth',shorten >=1pt,auto,node distance=2.8cm,semithick]
  \tikzstyle{every state}=[fill=none,draw=black,text=black]

  \node[state] (S) {$S$};
  \node[state] (B) [below left of=S] {$B$};
  \node[state] (F) [below right of=S] {$F$};

  \path (S) edge [bend right] node [left=5]{$f_i(t)$} (B)
  edge [bend left] node [right=5] {$g_i(t)$} (F)
  edge [loop above] node [above]{$1 - f_i(t) - g(t)$} (S)
  (B) edge [bend right] node [below]{$p_v\,(1-p_f)$} (F)
  edge [bend right] node [right] {$p_f$} (S)
  edge [loop left ] node [left] {$(1-p_v)\, (1-p_f)$} (B)
  (F) edge [bend left] node [right]{$p_f$} (S)
  edge [loop right ] node [right]{$(1-p_f)$} (F);
\end{tikzpicture}
	\caption{The transitions states  for the generic $i$-th agent of our hoax epidemic model. 
	To simplify the model, here we set $p_v = 1 - \alpha$.} 
	\label{fig:modelhoax}
\end{figure}
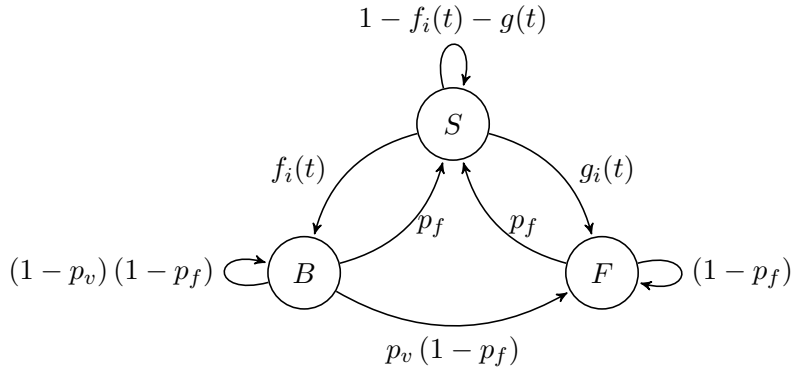

Here we describe a model of the spread of the belief in a \emph{hoax} and the related \emph{fact checking} within a social network of agents with finite memory. An agent can be  in any of the following three states: `Susceptible' 
(denoted by $S$), if they have not heard about neither the hoax nor the fact checking, or if they have forgotten about it; `Believer' ($B$), if they believe in the hoax and choose to spread it;  and `Fact checker' ($F$) if they know the hoax is false --- for example after having consulted an accurate news source --- and choose to spread the fact checking. 

Let us consider the $i$-th agent at time step $t$ and let us denote with $n_i^{X}(t)$ the number of its neighbors in state $X\in\left\{S,B,F\right\}$. 
We assume that an agent `decides' to believe in either the hoax or the 
fact checking as a result of interaction over interpersonal ties. This could be due to social conformity~\cite{asch1961effects} or because agents accept information from their neighbors~\cite{rosnow1976rumor}. Second, we assume that the hoax displays an intrinsic \emph{credibility} 
$\alpha\in \left[0,1\right]$, which, all else being equal, makes it more believable than the fact checking. We will discuss later how this parameter can be also related to the users: by now, we consider it as a feature of the hoax. Thus, the probability of transitioning from $S$ to either $B$ or $F$ are given by functions $f_i$, and $g_i$, respectively: 
\begin{align}
f_i(t) = \beta\,\frac{{n_i^B}(1 + \alpha)}{{n_i^B}(1 + \alpha) + {n_i^F}(1 - \alpha)}\\
g_i(t) = \beta\,\frac{{n_i^F}(1 - \alpha)}{{n_i^B}(1 + \alpha) + {n_i^F}(1 - \alpha)}
\label{eq:fg}
\end{align}
where $\beta \in [0,1]$ is the overall spreading rate. Furthermore, agents possess
finite memory, which means that at each time step, any believer of
fact checker can `forget' about the hoax with fixed probability $p_f$
or the fact check and become susceptible again. Finally, any believer who 
has not forgotten the hoax yet can decide to check the news and stop 
believing in the hoax, becoming a fact checker. This happens with 
probability $p_v$. In any other case, an agent remains in its current state. 
The full model with the transition states are shown in Fig.~\ref{fig:modelhoax}.

Observe that, since $f_i(t) + g_i(t)=\beta$, which is equivalent to the infection 
rate of the SIS model. Indeed, if one considers the two states $B$ and $F$ 
as single `Infected' state ($I$), then our model is reduced to an SIS model,
with the only difference that the probability of recovery $\mu$ is denoted by $p_f$.

Let us denote by $s_i(t)$ the state of the $i$-th agent at time $t$, and let us define, for $X \in \{B,F,S\}$, the state indicator function $s_i^X(t) = \delta(s_i(t), X)$.
The triple $p_i(t) = \left[p_i^{B}(t), p_i^{F}(t), p_i^{S}(t)\right]$ describes the  probability that a node $i$ is in any of the three states at time t. 
The dynamics of the system at time $t + 1$ will be then given by a random realization
of $p_i$ at $t + 1$. Thus, $p_i(t + 1)$ can be described as:
\begin{align}
{p_i^{B}(t+1)} &= f_i(t) s_i^{S}(t) + (1 - p_f)(1 - p_v) s_i^{B}(t)\\
{p_i^{F}(t+1)} &= g_i(t) s_i^{S}(t) + p_v (1 - p_f) s_i^{B}(t) + (1 - p_f) s_i^{F}(t)\\
{p_i^{S}(t+1)} &= p_f \left[s_i^{B}(t) + s_i^{F}(t)\right] + \left[1 - f_i(t) - g_i(t)\right] s_i^{S}(t)
\label{prop}
\end{align}

In previous work~\cite{tambuscio2015fact} we analyzed the behavior of the model at equilibrium. Starting from a well-mixed topology of $N$ agents, in which a few agents have been initially seeded as believers, we derived the expressions for the density of believers, fact checkers, and susceptible agents in the infinite-time limit denoted by $B_{\infty}$, $F_{\infty}$, and 
$S_{\infty}$, respectively. 
We found that, independent of the network topology (Barab\'asi-Albert and Erd\H{o}s-R\'enyi), the value of $p_v$, and of $\alpha$, $S_{\infty}$ stabilizes around the same values in all simulations. We confirmed such a result using both mean-field equations and simulations. 

Once the system reaches equilibrium, the relative ratio between believers and fact checkers is 
determined by  $\alpha$ and $p_v$: such as the higher $\alpha$, the more believers, 
and conversely for $p_v$. In particular, we showed that there always exists 
a critical value of $p_v$ above which the hoax is completely eradicated 
from the network (i.e., $B_\infty = 0$). This value depends on $\alpha$ 
and $p_f$, but not on the spreading rate $\beta$. 

As one can see, the model has several parameters, namely, spreading rate
$\beta$, credibility of the hoax $\alpha$, probability of verification
$p_v$ and probability of forgetting $p_f$. Since, in the present work, we want to consider the role of communities of people with different attitudes to believe to an hoax, the number of parameters is going to increase. 

As an attempt to reduce the number of parameters, we set
\begin{equation}
p_v = 1 - \alpha.
\label{eq:simplification}
\end{equation}

\begin{figure}[ht]
\centering 
\includegraphics[width=0.7\textwidth]{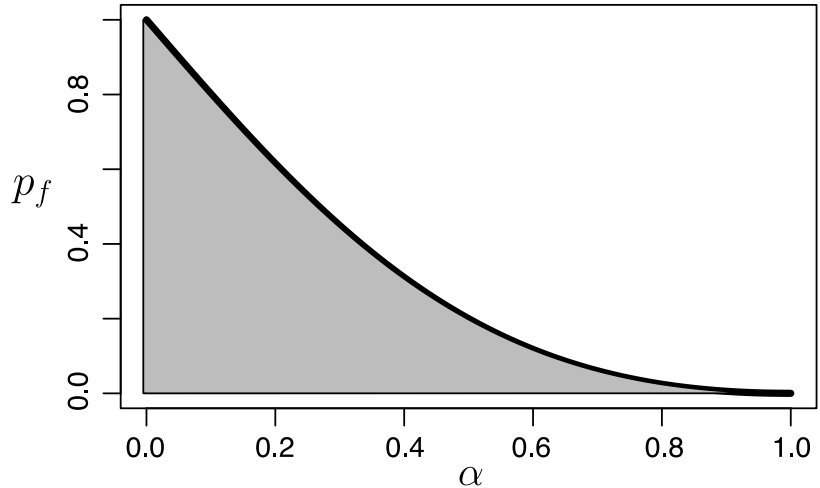}
\caption{
    Epidemic threshold for the simplified version of the model given by
    Eq.~\ref{eq:simplification}. The grey area indicates the region of the 
    parameter space where the hoax is completely removed from the network. 
    The white part indicates the region of the parameter space where the 
    hoax can become endemic.} 
\label{fig:mf2} 
\end{figure} 

\begin{figure}[ht]
	\centering
	\includegraphics[width=\textwidth]{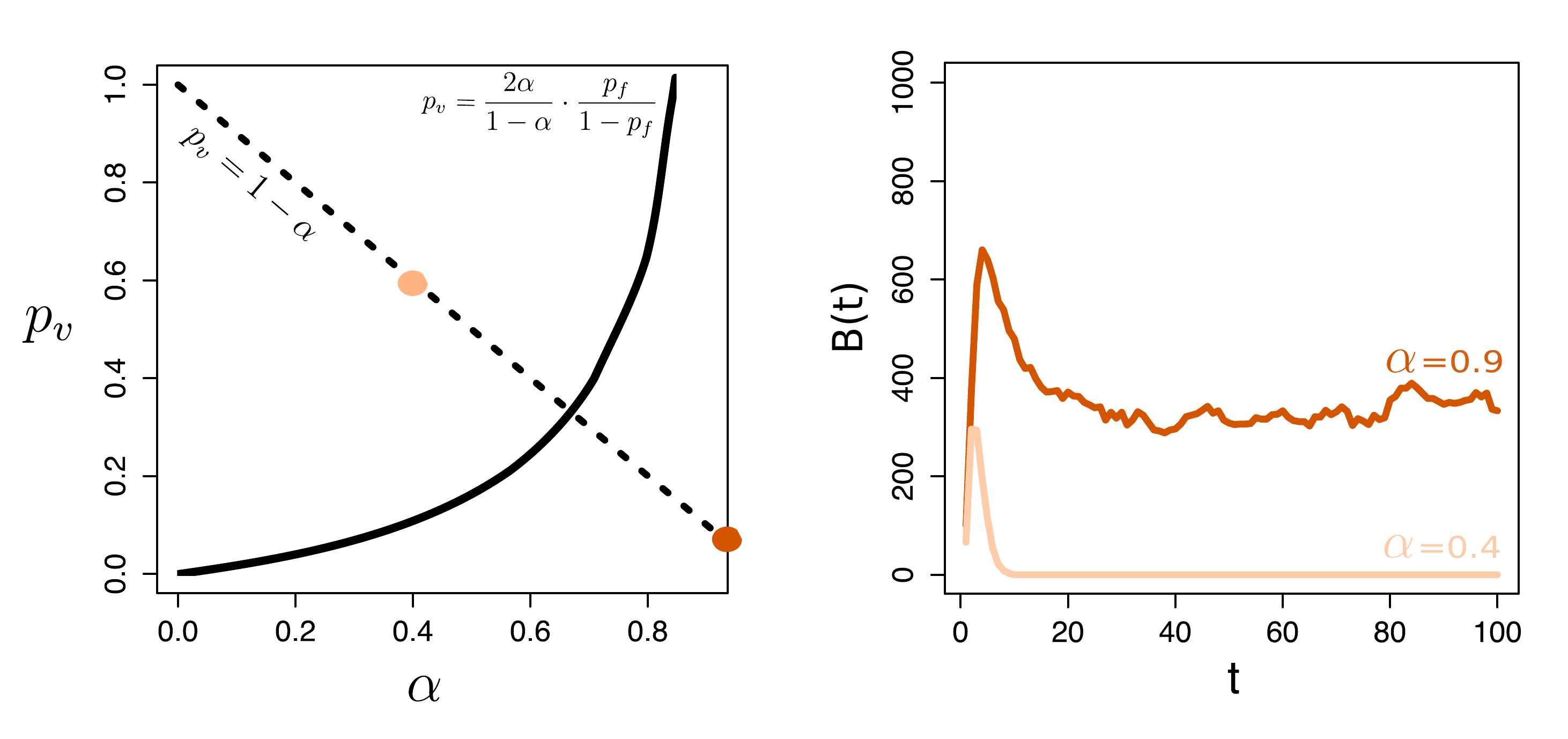} 
	\caption{Simplification of the model setting $p_v = 1- \alpha$, here fixing $p_f=0.1$. On the left we can observe the phase diagram of the entire parameter space considered for the model. In this new work we are restricting it to the dashed line, but we are keeping all the possible configurations of the model: believers can survive (dark line) or not (pale line), as we can see comparing the dots on the left to the curves on the right that show evolution of believers numbers in the network over time.}
	\label{fig:simplify}
\end{figure}

This simplification can be motivated by assuming that the more credible a piece of news is, the lower are the chances anybody will try to check its veracity. 
This means that we restrict the parameters space $p_v \times \alpha$ to a line. This constrain can be easily observed in Fig.~\ref{fig:simplify} (left), where the curve represents the analytic threshold on the verifying probability. Above it the hoax becomes endemic, over it the hoax is completely removed. We note that even with this additional constrain, this new, simplified model exhibits the same behaviors that our original model can produce (i.e., believers survive or not).

Recomputing the mean-field equations with Eq.~\ref{eq:simplification}, we obtain now a critical value for $p_f$, a sufficient condition that guarantees the removal of the hoax from the network: 
\begin{equation}
\label{eq:pfcond}
p_f \le \frac{(1 - \alpha)^2}{1 + \alpha^2} \quad\Longrightarrow\quad p^B(\infty) = 0.
\end{equation} 

The behaviour of $p_f$ versus $\alpha$ is shown in Fig.~\ref{fig:mf2}. For any combination of $p_f$ and $\alpha$ below the curve, the hoax is completely removed from the network. For combinations above the curve, the infection is instead endemic. The forgetting probability can be considered as a measure of the \emph{attention} toward a specific topic, meaning that if there is a large discussion around the subject, then exposed people tend to have a stable opinion about it, otherwise the probability to forget the belief and the news will be higher. The presence of this threshold in Eq.~\ref{eq:pfcond} could suggest that the level of attention plays an important role in fake news global spread and persistence.

\section{Results}

Two parameters govern the spreading dynamics in our model. These are the credibility $\alpha$ and forgetting probability $p_f$. To address our research question about the role of network structure and communities, we consider a simple generative model of a segregated network. Let us consider $N$ agents divided into two groups, one comprised by $t < N$ agents whose beliefs conform more to the hoax than the other one, which is comprised by the rest of the population. We call the former the \emph{gullible} group, while the latter the \emph{skeptic} group. To represent this in our framework, we set different values of $\alpha$ for each agent group (either $\agul$ or $\ask$, $\agul > \ask$). This is not a contradiction with what we said before: the credibility is a parameter describing the hoax, but of course is also related to the user attitude and personal worldviews, then it is reasonable to think to different values for different groups.  

\begin{figure}[t]
\centering
\includegraphics[width=\textwidth]{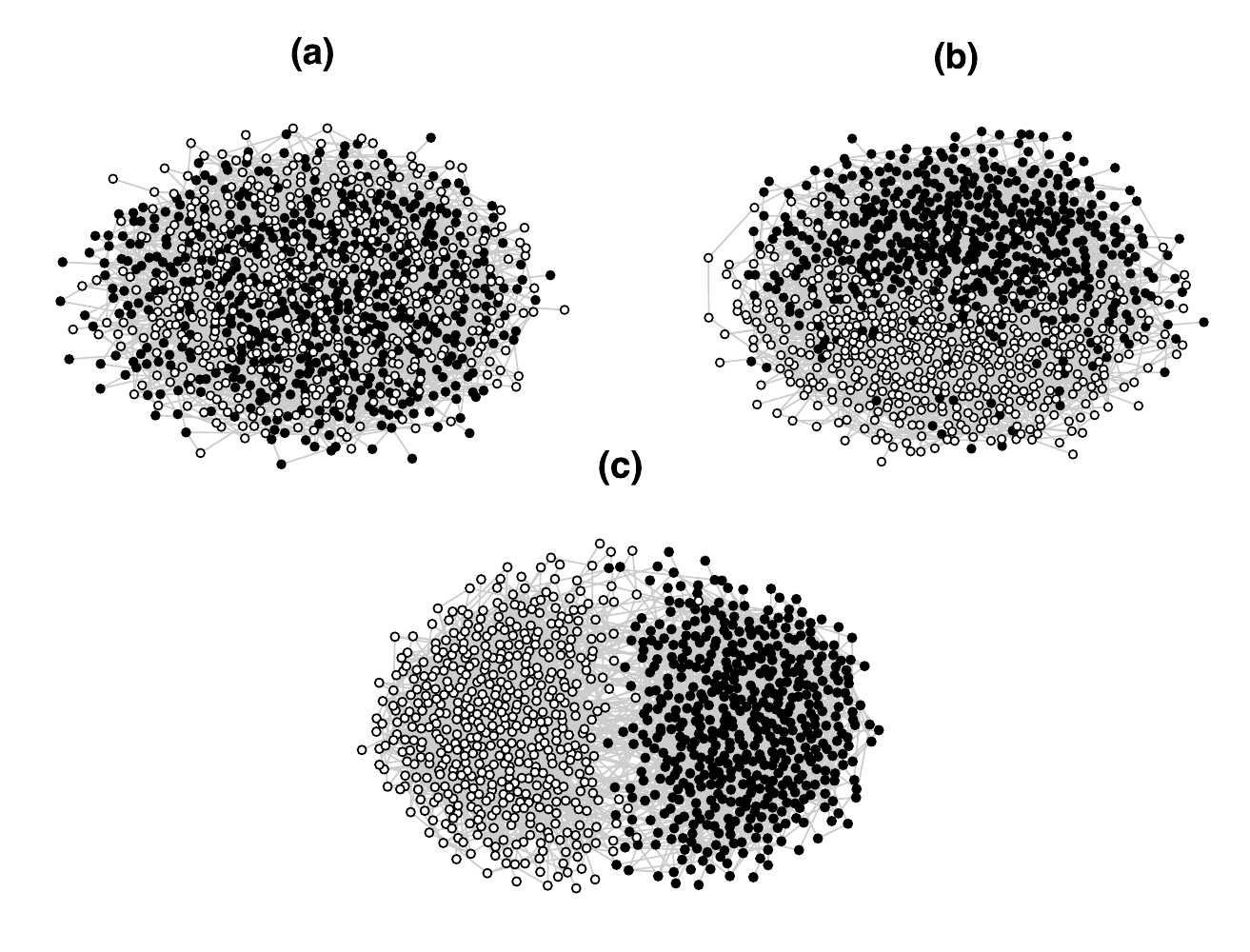} 
\caption{Network structure under different segregation regimes between 
    two groups (in this case of equal size). In the figure, three different values of $s$ were used. (\textbf{a}): 
    $s = 0.6$, (\textbf{b}): $s = 0.8$, and (\textbf{c}): $s = 0.95$. Node 
    layout was computed using a force-directed algorithm~\cite{Fruchterman1991}.}
\label{fig:segex}
\end{figure}

To generate the network, we assign $M$ edges at random. Let $s \in \left[\frac{1}{2}, 1\right)$ denote the fraction of intra-group edges,  regardless of the group. For each edge we first decide, with probability $s$, whether two individuals from the same group (intra-group tie) or different groups (inter-group tie) should be connected. In the case of an  intra-group tie, we select a group with probability proportional to the  relative ratio of the total number of possible inter-group ties (of that group) to that of the whole network; then, we pick uniformly at random two agents from that group and connect them. In the second case, two agents are chosen at random, one per group, and connected. Fig.~\ref{fig:segex} shows three examples of network with different values of $s$.

\begin{figure}[tbp]
\centering
\includegraphics[width=\textwidth]{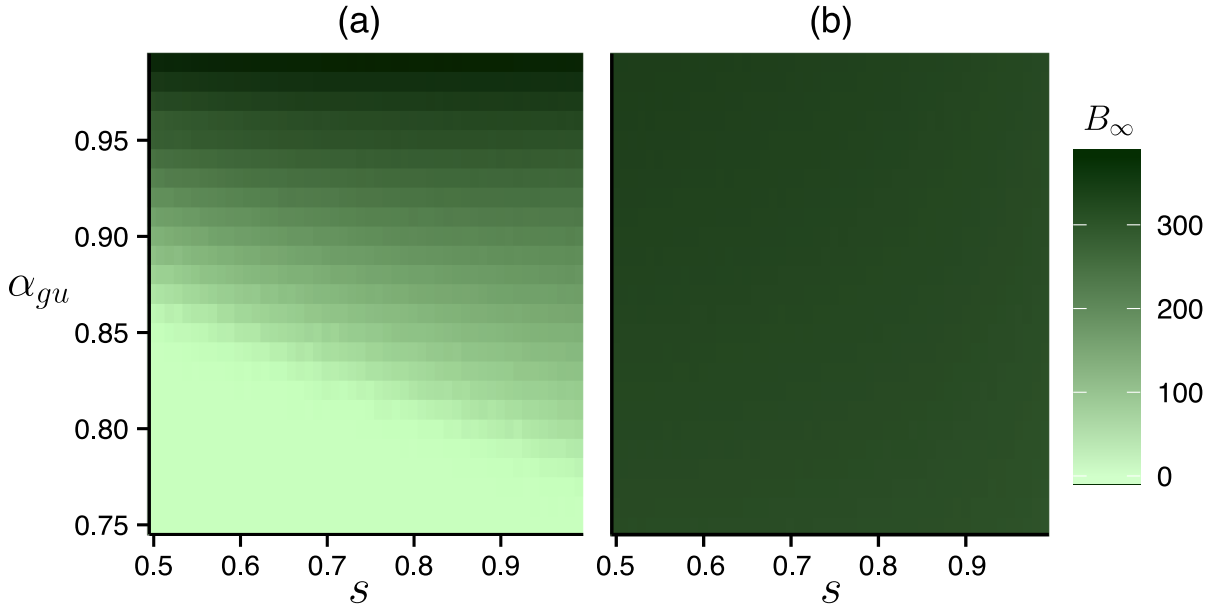}
\caption{Believers at equilibrium in the phase space of $s\times\agul$. We considered two forgetting regimes: (\textbf{a}): low forgetting, $p_f = 0.1$, and (\textbf{b}): high forgetting, $p_f = 0.8$. Other parameters: $\ask = 0.4$. Each point was averaged over 50 simulations.}
\label{fig:alphaseg}
\end{figure}

To understand the behavior of the model in this segregated network, 
we performed extensive numerical simulations. We set 
fixed values for $\ask$ and we considered a wide range of values of $\agul$, $p_f$, $s$, 
and $t$. Fig.~\ref{fig:alphaseg} reports the results of the first of 
these exercises, showing the overall number of believers $B_\infty$ in 
the whole population at equilibrium. 

Increasing either $\gamma$ or $\agul$ we see an increase of $B_\infty$, all else being equal. However, when we changed the segregation $s$ we observed two different situations: for small $p_f$, an increase of $s$ resulted in an increase of $B_{\infty}$. Conversely --- and perhaps a bit surprisingly --- under high values of $p_f$ increasing $s$ does not change $B_{\infty}$.  

\begin{figure}[th]
	\centering
	\includegraphics[width=.62\textwidth]{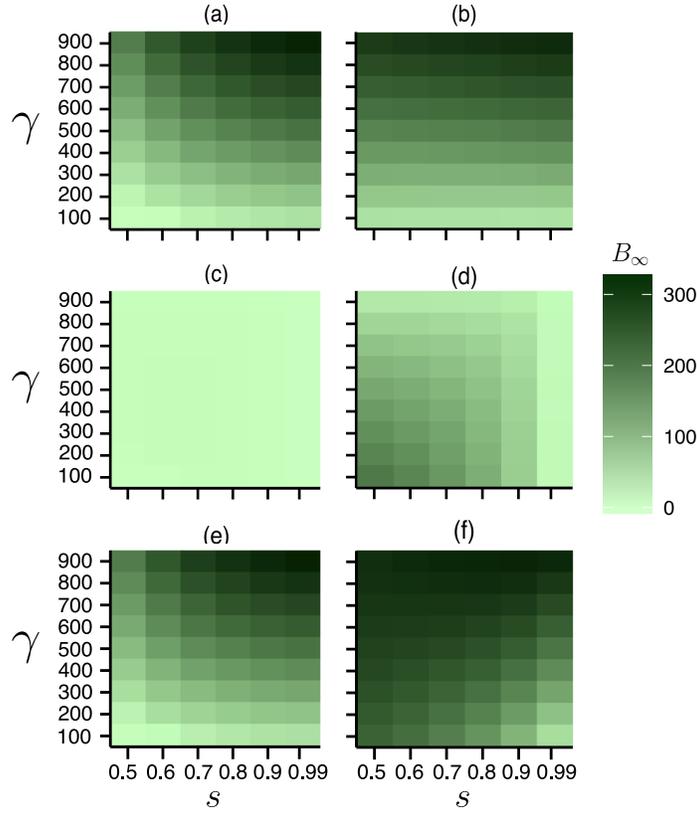} 
    \caption{Believers at equilibrium under low ($p_f = 0.1$) and high forgetting ($p_f = 0.8$) rate. The number of believers at equilibrium is broken down as $B_\infty = B_{\infty}^{\rm gu} + B_{\infty}^{\rm sk}$. 
    Phase diagrams in the space $s \times \gamma$ for (\textbf{a}) $B_\infty^{\rm gu}$, 
    low forgetting, (\textbf{b}) $B_\infty^{\rm gu}$, high forgetting, (\textbf{c}) 
    $B_\infty^{\rm sk}$ low forgetting, (\textbf{d}) $B_\infty^{\rm sk}$ 
    high forgetting, (\textbf{e}) $B_\infty$ low forgetting,
    and (\textbf{f}) $B_\infty$ high forgetting. We fixed $\agul = 0.9$ and 
    $\ask=0.05$.}
	\label{fig:segsize}
\end{figure}

Trying to better understand the role of $p_f$, we further explore the behavior of the model by varying the size of the gullible group $\gamma$ and its level of segregation $s$. In Fig.~\ref{fig:segsize} we report the relevant phase diagrams, breaking down the number of believers at equilibrium by group, i.e., $B_\infty = B_\infty^{\rm gu} + B_\infty^{\rm sk}$. If $p_f$ is low (Fig~\ref{fig:segsize}, left column), the overall number of believers depends heavily on $B_\infty^{\rm gu}$, whereas $B_\infty^{\rm sk} \approx 0$, and the segregation is unimportant,see Fig.~\ref{fig:segsize}(a) and Fig.~\ref{fig:segsize}(c). 

Instead, with an high rate of forgetting (right column), $B_{\infty}$ (Fig.~\ref{fig:segsize}(f)) depends on both $B_\infty^{\rm sk}$ and $B_\infty^{\rm gu}$. But in this case we have a different role of the segregation: while in the skeptic group $B_\infty^{\rm sk}$ decreases when $s$ increases, see Fig.~\ref{fig:segsize}(d), in the gullible group $s$ has fewer influence, see Fig.~\ref{fig:segsize}(b). 

\begin{figure}[tbp]
	\centering
	\includegraphics[width=0.7\textwidth]{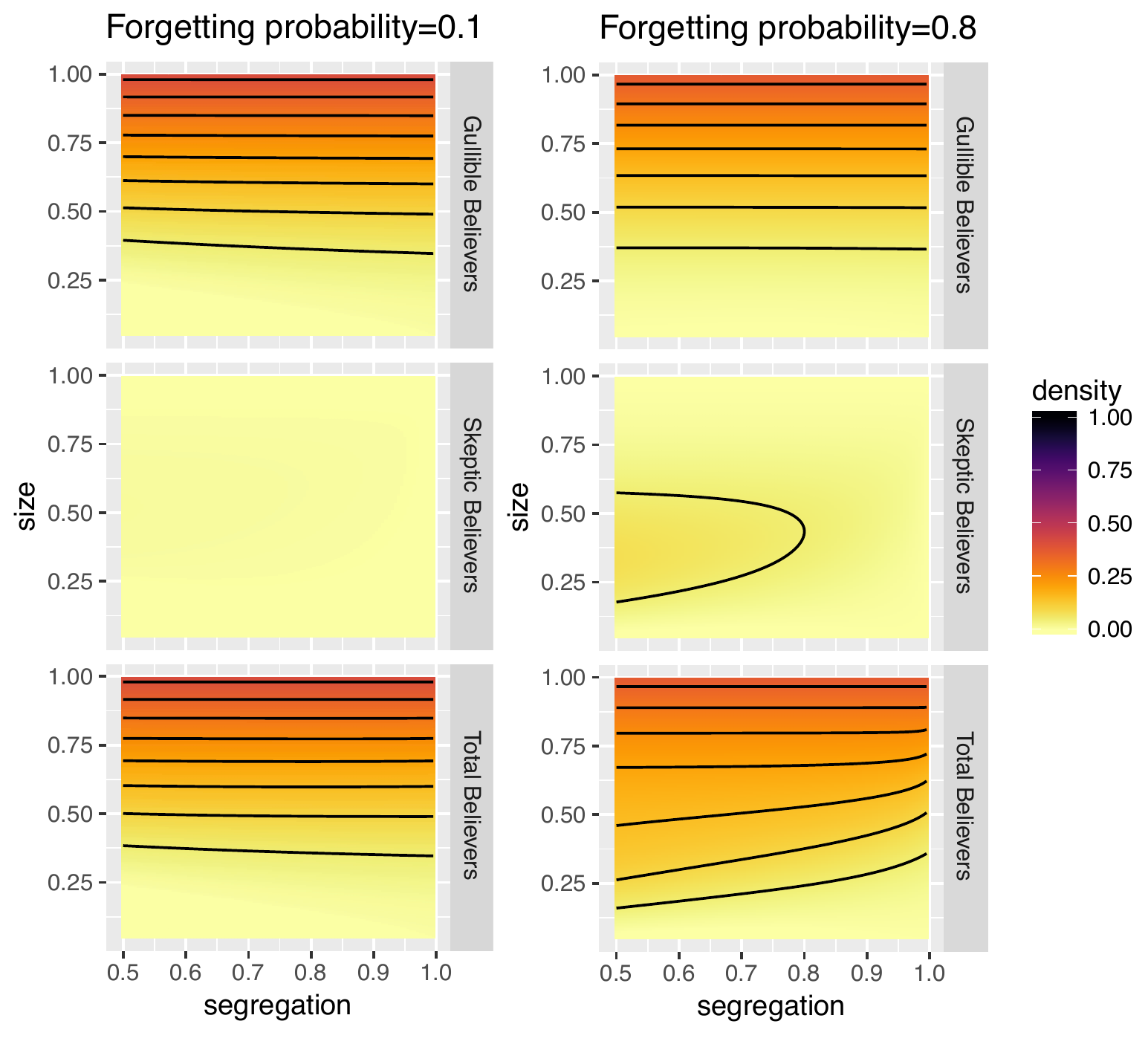} 
	\caption{Mean-field approximation for different values of $p_v$: these phase diagrams represent the density of Believers at equilibrium varying $\gamma$ and s, exactly as in Figure \ref{fig:segsize}.}
	\label{fig:mf_sim}
\end{figure}

To give an analytical support to our findings, we obtain mean-field approximation for the model (details in the Appendix) and we perform both numerical integration of the mean field equations and agent-based simulations, which give very similar results. Fig.~\ref{fig:mf_sim} shows the phase diagrams obtained by numerical simulations of the mean-field equations. 

Summarizing, segregation can have a very different role on the final configuration of the hoax spreading and this depends on the forgetting rate. Why the number of links among communities with different behaviors is so important? It should be noted that any `network effect' present in our model will only appear in the infection phase, that is for transitions $S\to B$ and $S\to F$. To better understand what happens in both groups, we computed the rate at which these transitions happen, that is, the conditional probability of, being susceptible, becoming either believer or fact checker. 

\begin{figure}[ht]
\centering
\includegraphics[width=\textwidth]{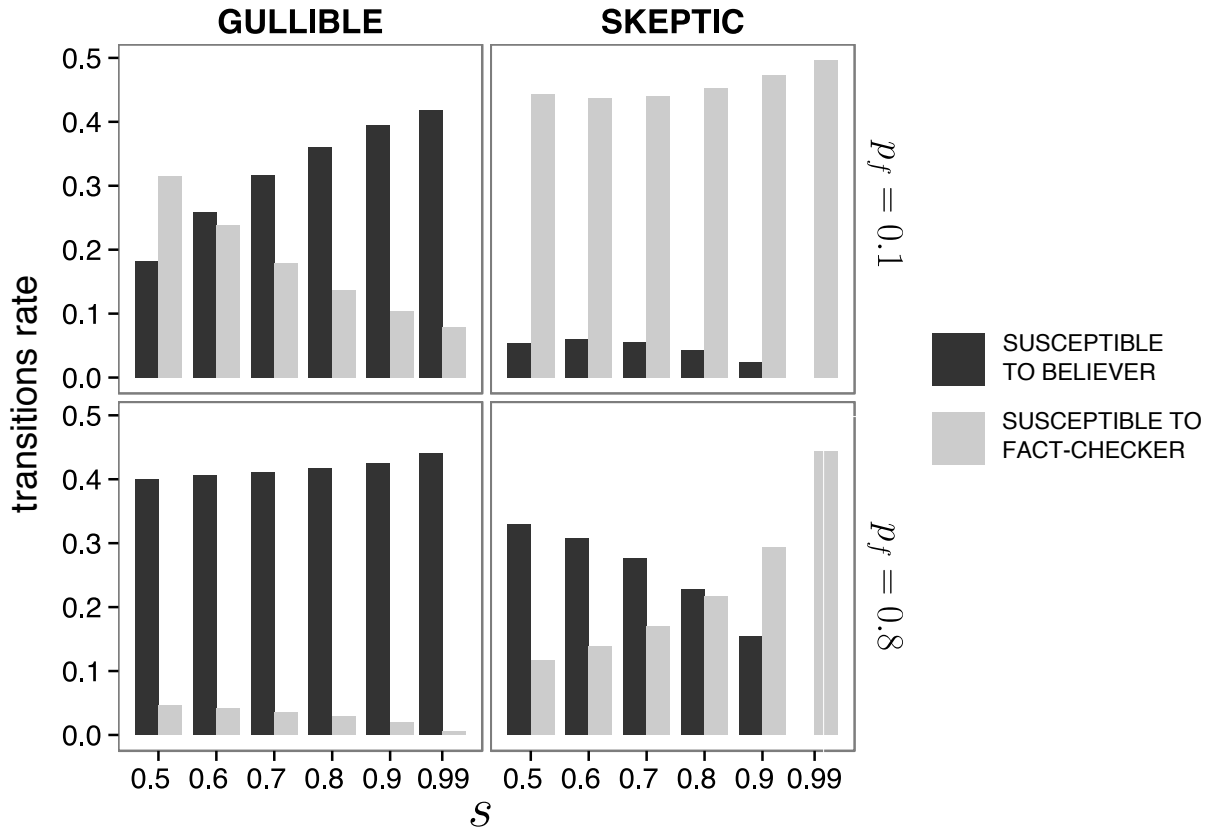} 
\caption{Rate of transitions of type $S\to B$ and $S\to F$ at equilibrium. We run the simulation until the system has reached the steady state and then compute the average number of transitions per susceptible. The plot shows averages over 50 simulations.}
\label{fig:histotrans}
\end{figure}

Let us consider a susceptible agent in the gullible group. At low forgetting rates, in the gullible group more intra-group ties (i.e.~an higher $s$) increase the chances of becoming a believer and reduce those of becoming fact checker; see Fig.~\ref{fig:histotrans} (top left). In the skeptic group, the segregation effect is almost negligible (top right). This happens because inter-group ties expose the susceptible agents, among the gullible, to more members of the skeptic group, who are largely fact checkers. 

At high forgetting rates, instead, we observe the opposite behavior: more inter-group ties translate into more exposure, for susceptible users in the skeptic group, to gullible agents who are, by and large, believers. In the gullible group (bottom left of Fig.~\ref{fig:histotrans}), segregation is not very important while, in the skeptic group, more connections with the gullible mean more believers (bottom right of Fig.~\ref{fig:histotrans}).

In other words, the role of segregation, being related to the abundance of inter-group ties, can have both a positive and negative role in stopping the spread of misinformation: for low forgetting rates, these links can help the spread of the debunking in the gullible group, while for high forgetting rates, they have the opposite effect, helping the hoax spread in the skeptic group. 

\section{Discussion}

Using agent-based simulations, here we have analyzed the role of the underlying structure of the network on which the diffusion of a piece of misinformation takes place. In particular we consider a network formed by two groups --- gullible and skeptic --- characterized by different values of the credibility parameter $\alpha$. In order to study how the social structure shapes information exposure, we introduce a parameter $s$ that regulates the abundance of ties between these two groups. We observe that $s$ has an important role in the diffusion of misinformation. If the probability of forgetting $p_f$ is small then the fraction of the population affected by the hoax will be large or small depending on whether the network is respectively segregated or not. However, if the rate of forgetting is large, segregation has no effect on the spread of the hoax.

The probability of forgetting could be also interpreted as a measure of how much a given topic is discussed. A low value of $p_f$ could perhaps fit well with the scenario of a ideas whose belief tends to be more persistent over time, for example conspiracy theories. A high value $p_f$ could fit better with situations where beliefs are short lived, either because the claims are easy to debunk or are no more interesting than mere gossip, whose information value is transient. Hoaxes about the alleged death of celebrities, for instance, could fall within this latter category. 

On the basis of the findings presented in this paper, further research should be devoted to understanding the role of segregation in the spread of misinformation on social media. For conspiracy theories, it could be useful to analyze what happens if the communication among different groups increases. Moreover, it could be also interesting to consider more realistic situations in which  credibility is distributed over the population according to socio-economic features and other individual-level attributes.  

Our results are also important from a purely theoretical point of view. Indeed, the model we had introduced in prior work, and on which we have build upon here, was an example of an epidemic process that is not affected by the network topology, meaning that the structure does not influence the final configuration of the network --- indeed, it could be proved that there are no significant differences in the behavior of the spreading dynamics in random or scale-free networks~\cite{tambuscio2015fact}. In the present work, however, we show that actually network structure can become very important if we add a layer of complexity, characterizing groups of nodes with slightly different behaviors (here different values of the credibility parameter). This can suggest a lot of research, experiments and simulations in order to understand which parameters are sensitive to the segregation level, or some other topology measure, even in models that have usually topology-independent dynamics.

In conclusion, understanding the production and consumption of misinformation is a critical issue~\cite{Ciampaglia2015a}. As several episodes are showing, there are obvious consequences connected to the uncontrolled production and consumption of inaccurate information~\cite{howell2013digital}. A more thorough understanding of rumor propagation and the structural properties of the information exchange networks on which this happens may help mitigate these risks.

\section*{Acknowledgments}
The authors would like to acknowledge Filippo Menczer and Alessandro Flammini for feedback and insightful conversations. DFMO acknowledges the support from James S. McDonnell Foundation.  GLC acknowledges support from the Indiana University Network Science Institute (\url{iuni.iu.edu}) and from the Swiss National Science Foundation (PBTIP2\_142353).

\appendix

\section{Appendix}

\subsection{Mean-Field computations}

In previous work we showed mean-field analysis for our model on a homogeneous network~\cite{tambuscio2015fact}. Similarly, here we perform a similar analysis for our model on a network segregated in two groups, skeptic and gullible: for each group we have three equations (see Eq.~\ref{prop}) representing the spreading process. 

\begin{align*}
{p_{i_g}^{B}(t+1)} &= f_{i_g}(t) s_{i_g}^{S}(t) + (1 - p_f)(1 - p_v) s_{i_g}^{B}(t)\\
{p_{i_g}^{F}(t+1)} &= g_{i_g}(t) s_{i_g}^{S}(t) + p_v (1 - p_f) s_{i_g}^{B}(t) + (1 - p_f) s_{i_g}^{F}(t)\\
{p_{i_g}^{S}(t+1)} &= p_f \left[s_{i_g}^{B}(t) + s_{i_g}^{F}(t)\right] + \left[1 - f_{i_g}(t) - g_{i_g}(t)\right] s_{i_g}^{S}(t)\\[2ex]
{p_{i_{sk}}^{B}(t+1)} &= f_{i_{sk}}(t) s_{i_{sk}}^{S}(t) + (1 - p_f)(1 - p_v) s_{i_{sk}}^{B}(t)\\
{p_{i_{sk}}^{F}(t+1)} &= g_{i_{sk}}(t) s_{i_{sk}}^{S}(t) + p_v (1 - p_f) s_{i_{sk}}^{B}(t) + (1 - p_f) s_{i_{sk}}^{F}(t)\\
{p_{i_{sk}}^{S}(t+1)} &= p_f \left[s_{i_{sk}}^{B}(t) + s_{i_{sk}}^{F}(t)\right] + \left[1 - f_{i_{sk}}(t) - g_{i_{sk}}(t)\right] s_{i_{sk}}^{S}(t)
\end{align*}

In these equations we can substitute $s_i(t)$ with $p_i(t)$ and when $t \to \infty $ we can assume $p_i(t)=p_i(t+1)=p_i(\infty)$ for all $i \in N$. Hereafter we simplify the notation using $p_g^B(\infty)=p_g^B$ (and analogously for the other cases). Now, let us consider the spreading functions for the gullible agents. Similar equations can be written for the case of skeptic agents. The spreading functions are: 

\begin{align*}
f_{i_g}(t) = \beta\,\frac{{n_{i_g}^B}(1 + \alpha)}{{n_{i_g}^B}(1 + \alpha) + {n_{i_g}i^F}(1 - \alpha)}\\
g_{i_g}(t) = \beta\,\frac{{n_{i_g}^F}(1 - \alpha)}{{n_{i_g}^B}(1 + \alpha) + {n_{i_g}^F}(1 - \alpha)}
\end{align*}

Assuming that all vertices have the same number of neighbors $\langle k \rangle$, and that these neighbors are chosen randomly, we can write $n_i^B= s\cdot n^B{i_g} + (1-s) \cdot n^B_{i_{sk}}$, where $n^B_{i_g}= \gamma \cdot \langle k \rangle p^B_g$ and $n^B_{i_{sk}}= (1-\gamma) \cdot \langle k \rangle p^B_{sk}$. Similarly, for $n_i^F$, we can obtain an expression that is not dependent on $i$. This simplifies the equations and lets us to simulate the process iterating the application of them until the values of $p^S_{sk},p^B_{sk}, p^F_{sk}, p^S_{g}, p^B_{g},p^F_{g}$ have reached stability.


\begin{thebibliography}{10}

\bibitem{acemoglu2010spread}
D.~Acemoglu, A.~Ozdaglar, and A.~ParandehGheibi.
\newblock Spread of (mis) information in social networks.
\newblock {\em Games and Economic Behavior}, 70(2):194--227, 2010.

\bibitem{allport1947psychology}
G.~W. Allport and L.~Postman.
\newblock {\em The psychology of rumor.}
\newblock Oxford, England: Henry Holt, 1947.

\bibitem{anagnostopoulos2014viral}
A.~Anagnostopoulos, A.~Bessi, G.~Caldarelli, M.~Del~Vicario, F.~Petroni,
  A.~Scala, F.~Zollo, and W.~Quattrociocchi.
\newblock Viral misinformation: the role of homophily and polarization.
\newblock {\em arXiv preprint arXiv:1411.2893}, 2014.

\bibitem{Andrews:2016:KUT:2818048.2819986}
C.~Andrews, E.~Fichet, Y.~Ding, E.~S. Spiro, and K.~Starbird.
\newblock Keeping up with the tweet-dashians: The impact of 'official' accounts
  on online rumoring.
\newblock In {\em Proceedings of the 19\textsuperscript{th} ACM Conference on
  Computer-Supported Cooperative Work \& Social Computing}, CSCW '16, pages
  452--465, New York, NY, USA, 2016. ACM.

\bibitem{asch1961effects}
S.~E. Asch.
\newblock Effects of group pressure upon the modification and distortion of
  judgements.
\newblock In M.~Henle, editor, {\em Documents of Gestalt Psychology}, pages
  222--236. University of California Press, Oakland, California, USA, 1961.

\bibitem{Bakshy2011}
E.~Bakshy, J.~M. Hofman, W.~A. Mason, and D.~J. Watts.
\newblock Everyone's an influencer: Quantifying influence on twitter.
\newblock In {\em Proceedings of the Fourth ACM International Conference on Web
  Search and Data Mining}, WSDM '11, pages 65--74, New York, NY, USA, 2011.
  ACM.

\bibitem{Bakshy2015}
E.~Bakshy, S.~Messing, and L.~A. Adamic.
\newblock Exposure to ideologically diverse news and opinion on facebook.
\newblock {\em Science}, 348(6239):1130--1132, 2015.

\bibitem{bakshy2012role}
E.~Bakshy, I.~Rosenn, C.~Marlow, and L.~Adamic.
\newblock The role of social networks in information diffusion.
\newblock In {\em Proceedings of the 21\textsuperscript{st} international
  conference on World Wide Web}, pages 519--528. ACM, 2012.

\bibitem{benkler2006wealth}
Y.~Benkler.
\newblock {\em The wealth of networks: How social production transforms markets
  and freedom}.
\newblock Yale University Press, 2006.

\bibitem{Bessi2015}
A.~Bessi, M.~Coletto, G.~A. Davidescu, A.~Scala, G.~Caldarelli, and
  W.~Quattrociocchi.
\newblock Science vs conspiracy: Collective narratives in the age of
  misinformation.
\newblock {\em PLoS ONE}, 10(2):e0118093, Feb. 2015.

\bibitem{borel2016chicago}
B.~Borel.
\newblock {\em The Chicago Guide to Fact-Checking}.
\newblock The University of Chicago Press, Chicago, IL, USA, 2016.

\bibitem{butler2011hypercorrection}
A.~C. Butler, L.~K. Fazio, and E.~J. Marsh.
\newblock The hypercorrection effect persists over a week, but high-confidence
  errors return.
\newblock {\em Psychonomic bulletin \&amp; review}, 18(6):1238--1244, 2011.

\bibitem{centola2007complex}
D.~Centola and M.~Macy.
\newblock Complex contagions and the weakness of long ties.
\newblock {\em American journal of Sociology}, 113(3):702--734, 2007.

\bibitem{chierichetti2009rumor}
F.~Chierichetti, S.~Lattanzi, and A.~Panconesi.
\newblock Rumor spreading in social networks.
\newblock In {\em Automata, Languages and Programming}, pages 375--386.
  Springer, 2009.

\bibitem{Ciampaglia2015a}
G.~L. Ciampaglia, A.~Flammini, and F.~Menczer.
\newblock The production of information in the attention economy.
\newblock {\em Scientific Reports}, 5:9452, 2015.

\bibitem{Ciampaglia2015}
G.~L. Ciampaglia, P.~Shiralkar, L.~M. Rocha, J.~Bollen, F.~Menczer, and
  A.~Flammini.
\newblock Computational fact checking from knowledge networks.
\newblock {\em PLoS ONE}, 10(6):e0128193, June 2015.

\bibitem{daley1964epidemics}
D.~J. Daley and D.~G. Kendall.
\newblock Epidemics and rumours.
\newblock {\em Nature}, 204:1118, 1964.

\bibitem{DelVicario2016}
M.~Del~Vicario, A.~Bessi, F.~Zollo, F.~Petroni, A.~Scala, G.~Caldarelli, H.~E.
  Stanley, and W.~Quattrociocchi.
\newblock The spreading of misinformation online.
\newblock {\em Proceedings of the National Academy of Sciences}, 2016.

\bibitem{dong2015knowledge}
X.~L. Dong, E.~Gabrilovich, K.~Murphy, V.~Dang, W.~Horn, C.~Lugaresi, S.~Sun,
  and W.~Zhang.
\newblock Knowledge-based trust: Estimating the trustworthiness of web sources.
\newblock {\em Proceedings of the VLDB Endowment}, 8(9):938--949, 2015.

\bibitem{Factcheck.org}
Factcheck.org.
\newblock A project of the annenberg public policy center, Oct. 2017.
\newblock [Online; Last accessed: 28-Oct-2017].

\bibitem{friggeri2014rumor}
A.~Friggeri, L.~A. Adamic, D.~Eckles, and J.~Cheng.
\newblock Rumor cascades.
\newblock In {\em Proc. Eighth Intl. AAAI Conf. on Weblogs and Social Media
  (ICWSM)}, pages 101--110, 2014.

\bibitem{Fruchterman1991}
T.~M.~J. Fruchterman and E.~M. Reingold.
\newblock Graph drawing by force-directed placement.
\newblock {\em Software: Practice \& Experience}, 21(11):1129--1164, Nov. 1991.

\bibitem{funk2010interacting}
S.~Funk and V.~A.~A. Jansen.
\newblock Interacting epidemics on overlay networks.
\newblock {\em Phys. Rev. E}, 81:036118, Mar. 2010.

\bibitem{galam2003modelling}
S.~Galam.
\newblock Modelling rumors: the no plane pentagon french hoax case.
\newblock {\em Physica A: Statistical Mechanics and Its Applications},
  320:571--580, 2003.

\bibitem{gleeson2014competition}
J.~P. Gleeson, J.~A. Ward, K.~P. O'Sullivan, and W.~T. Lee.
\newblock Competition-induced criticality in a model of meme popularity.
\newblock {\em Phys. Rev. Lett.}, 112:048701, Jan. 2014.

\bibitem{howell2013digital}
L.~Howell et~al.
\newblock Digital wildfires in a hyperconnected world.
\newblock In {\em Global Risks}. World Economic Forum, 2013.

\bibitem{knapp1944psychology}
R.~H. Knapp.
\newblock A psychology of rumor.
\newblock {\em Public opinion quarterly}, 8(1):22--37, 1944.

\bibitem{Kwak:2010:TSN:1772690.1772751}
H.~Kwak, C.~Lee, H.~Park, and S.~Moon.
\newblock What is twitter, a social network or a news media?
\newblock In {\em Proceedings of the 19\textsuperscript{th} International
  Conference on World Wide Web}, WWW '10, pages 591--600, New York, NY, USA,
  2010. ACM.

\bibitem{McPherson2001}
M.~McPherson, L.~Smith-Lovin, and J.~M. Cook.
\newblock Birds of a feather: Homophily in social networks.
\newblock {\em Annual Review of Sociology}, 27(1):415--444, 2001.

\bibitem{moreno2004dynamics}
Y.~Moreno, M.~Nekovee, and A.~F. Pacheco.
\newblock Dynamics of rumor spreading in complex networks.
\newblock {\em Physical Review E}, 69(6):066130, 2004.

\bibitem{nematzadeh2017how}
A.~Nematzadeh, G.~L. Ciampaglia, F.~Menczer, and A.~Flammini.
\newblock How algorithmic popularity bias hinders or promotes quality.
\newblock e-print, CoRR, July 2017.

\bibitem{newman2013interacting}
M.~E.~J. Newman and C.~R. Ferrario.
\newblock Interacting epidemics and coinfection on contact networks.
\newblock {\em PLOS ONE}, 8(8):1--8, Aug. 2013.

\bibitem{nikolov2015measuring}
D.~Nikolov, D.~F. Oliveira, A.~Flammini, and F.~Menczer.
\newblock Measuring online social bubbles.
\newblock {\em PeerJ Computer Science}, 1:e38, 2015.

\bibitem{Note1}
The Duke Reporters' Lab keeps an updated list of global fact-checking sites at
  \protect \url {https://reporterslab.org/fact-checking/}.

\bibitem{nyhan2015effect}
B.~Nyhan and J.~Reifler.
\newblock The effect of fact-checking on elites: A field experiment on us state
  legislators.
\newblock {\em American Journal of Political Science}, 59(3):628--640, 2015.

\bibitem{Nyhan2013}
B.~Nyhan, J.~Reifler, and P.~A. Ubel.
\newblock The hazards of correcting myths about health care reform.
\newblock {\em Medical Care}, 51(2):127--132, 2013.

\bibitem{onnela2007structure}
J.-P. Onnela, J.~Saram{\"a}ki, J.~Hyv{\"o}nen, G.~Szab{\'o}, D.~Lazer,
  K.~Kaski, J.~Kert{\'e}sz, and A.-L. Barab{\'a}si.
\newblock Structure and tie strengths in mobile communication networks.
\newblock {\em Proceedings of the national academy of sciences},
  104(18):7332--7336, 2007.

\bibitem{Owens2015}
E.~Owens and U.~Weinsberg.
\newblock News feed fyi: Showing fewer hoaxes, 2015.
\newblock [Online; accessed January 2016].

\bibitem{Pariser2011}
E.~Pariser.
\newblock {\em The filter bubble: What the Internet is hiding from you}.
\newblock Penguin UK, 2011.

\bibitem{Pastor-Satorras2015}
R.~Pastor-Satorras, C.~Castellano, P.~Van~Mieghem, and A.~Vespignani.
\newblock Epidemic processes in complex networks.
\newblock {\em Rev. Mod. Phys.}, 87:925--979, Aug. 2015.

\bibitem{qiu2017limited}
X.~Qiu, D.~F. Oliveira, A.~S. Shirazi, A.~Flammini, and F.~Menczer.
\newblock Limited individual attention and online virality of low-quality
  information.
\newblock {\em Nature Human Behaviour}, 1(7):s41562--017, 2017.

\bibitem{rosnow1976rumor}
R.~L. Rosnow and G.~A. Fine.
\newblock {\em Rumor and gossip: The social psychology of hearsay.}
\newblock Elsevier, 1976.

\bibitem{Snopes.com}
Snopes.com.
\newblock The definitive fact-checking site and reference source for urban
  legends, folklore, myths, rumors, and misinformation., Oct. 2017.
\newblock [Online; last accessed: 28-Oct-2017.

\bibitem{sunstein2002republiccom}
C.~Sunstein.
\newblock {\em Republic.com}.
\newblock Princeton University Press, 2002.

\bibitem{tambuscio2015fact}
M.~Tambuscio, G.~Ruffo, A.~Flammini, and F.~Menczer.
\newblock Fact-checking effect on viral hoaxes: A model of misinformation
  spread in social networks.
\newblock In {\em Proceedings of the 24\textsuperscript{th} International
  Conference on World Wide Web Companion}, pages 977--982. International World
  Wide Web Conferences Steering Committee, 2015.

\bibitem{Politifact.com}
T.~B. Times.
\newblock Fact-checking u.s. politics, Oct. 2017.
\newblock [Online; last accessed: 28-Oct-2017].

\bibitem{weng2012competition}
L.~Weng, A.~Flammini, A.~Vespignani, and F.~Menczer.
\newblock Competition among memes in a world with limited attention.
\newblock {\em Scientific Reports}, 2, 2012.

\bibitem{wood2016elusive}
T.~Wood and E.~Porter.
\newblock The elusive backfire effect: Mass attitudes' steadfast factual
  adherence.
\newblock e-print, SSRN, Aug. 2016.

\end{thebibliography}
\end{document}